\documentclass{article}
\usepackage{amssymb}
\usepackage{amsfonts}
\usepackage{amsmath}
\usepackage{graphicx}
\usepackage{cite}
\usepackage{hyperref}

\begin{document}

\title{Effective equation for two coupled oscillators: towards a global view of metamorphoses of the
amplitude profiles}
\author{Jan Kyzio\l , Andrzej Okni\'nski \\
Politechnika \'Swi\c{e}tokrzyska, Al. 1000-lecia PP 7, \\
25-314 Kielce, Poland}
\maketitle

\begin{abstract}
Dynamics of nonlinear coupled driven oscillators is investigated. Recently, we have
demonstrated that the amplitude profiles -- dependence of 
the amplitude $A$ on frequency $\Omega$ of the driving force, 
computed by asymptotic methods in implicit form as $F\left( A,\Omega \right) =0$, 
permit prediction of
metamorphoses of dynamics which occur at singular points of 
the implicit curve $F\left( A,\Omega \right) =0$.
In the present study we strive at a global view of singular points of the
amplitude profiles computing bifurcation sets, i.e. sets containing all points in the parameter 
space for which the amplitude profile has a singular point. 
\end{abstract}

\section{Introduction}
\label{Intro}

\begin{figure}[h!]
\center 
\includegraphics[width=10cm, height=4cm]{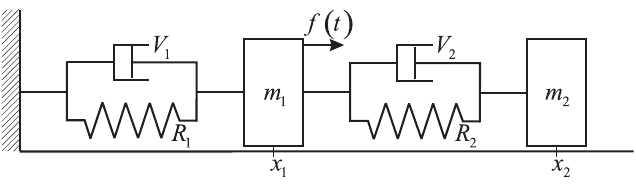}
\caption{Dynamic vibration absorber. Definitions of $R_1$, $V_1$, $R_2$, $V_2$ adopted in this work are given
in Eqs. (\ref{R1V1}), (\ref{R2V2}.}
\label{FF1}
\end{figure}

In this work we continue our investigation of coupled oscillators, see \cite%
{Kyziol2017} and references therein. Coupled oscillators play important role
in many scientific fields, e.g. neuroscience, chemistry, electronics, and
mechanics, see \cite%
{Mahmoud2004,Pikovsky2015,Schultheiss2011,Awal2019,Hajjaj2019,
Kozlowski1995,Kuznetsov2009,Perkins2012,Sabrathinam2013,Zulli2016,Luo2017,Karahan2017,Papangelo2019}
and references therein. A classic example is a dynamic vibration absorber,
consisting of a mass $m_{2}$, attached to the primary vibrating system of
mass $m_{1}$ \cite{DenHartog1985,Oueini1999} and described by equations:%
\begin{equation}
\left. 
\begin{array}{l}
m_{1}\ddot{x}_{1}-V_{1}\left( \dot{x}_{1}\right) -R_{1}\left( x_{1}\right)
+V_{2}\left( \dot{x}_{2}-\dot{x}_{1}\right) +R_{2}\left( x_{2}-x_{1}\right)
=f\left( t\right)  \\ 
m_{2}\ddot{x}_{2}-V_{2}\left( \dot{x}_{2}-\dot{x}_{1}\right) -R_{2}\left(
x_{2}-x_{1}\right) =0%
\end{array}%
\right\}   \label{gen-model}
\end{equation}%
where $f\left( t\right) =f\cos \left( \omega t\right) $, $V_{1}$, $R_{1}$
and $V_{2}$, $R_{2}$ represent (nonlinear) force of internal friction and
(nonlinear) elastic restoring force for mass $m_{1}$ and mass $m_{2}$,
respectively, see Fig. \ref{FF1}.

After making a a simplifying assumption:%
\begin{equation}
R_{1}\left( x_{1}\right) =-\alpha _{1}x_{1},\ V_{1}\left( \dot{x}_{1}\right)
=-\nu _{1}\dot{x}_{1},  \label{R1V1}
\end{equation}%
we derived approximate second-order effective equation and computed
nonlinear resonances $A\left( \omega \right) \cos \left( \omega t+\varphi
\right) $ in implicit form $F\left( A,\omega ;c_{1},c_{2},\ldots \right) =0$
applying the Krylov-Bogoliubov-Mitropolsky (KBM) method \cite{Nayfeh2011}
where $c_{1},c_{2},\ldots $ are parameters \cite{Okninski2006}. As explained 
in Section \ref{General} and demonstrated in our earlier papers 
bifurcations occur at singular points of the amplitude equation 
$F\left(A,\omega \right) =0$. In this work we attempt to find a global picture of
singular points of the amplitude equation in the case of the effective
equation, extending our results obtained for the van der Pol-Duffing
equation \cite{Kyziol2019}.

The paper is organized as follows. In the next Section the effective
equation and the amplitude profile of the nonlinear resonance, obtained in
Ref. \cite{Okninski2006}, are described. In Section \ref{General} singular points of 
the amplitude equation are defined and their relation to bifurcation of dynamics is explained. 
In Section \ref{Global} theory of algebraic curves is applied to compute the bifurcation 
set for the effective equation obtaining thus a global view of bifurcations. 
Plots of the bifurcation set and examples of singular points as well as corresponding bifurcations 
are presented in Section \ref{Numerical} and we summarize our results in the last Section.

\section{Approximate effective equation and the
Krylov-Bogoliubov-Mitropolsky amplitude profile}
\label{effective}

In new variables, $x\equiv x_{1}$, $y\equiv x_{2}-x_{1}$, we can eliminate
variable $x$ to obtain the following exact equation for relative motion \cite%
{Okninski2006,Kyziol2013b}: 
\begin{equation}
\left( M\frac{d^{2}}{dt^{2}}+\nu _{1}\frac{d}{dt}+\alpha _{1}\right) \left(
\mu \ddot{y}-V_{2}\left( \dot{y}\right) -R_{2}\left( y\right) \right)
+\epsilon m_{2}\left( \nu _{1}\frac{d}{dt}+\alpha _{1}\right) \ddot{y}=\hat{f%
}\left( t\right) ,  \label{4th}
\end{equation}%
where $\hat{f}\left( t\right) =m_{2}\omega ^{2}f\cos \left( \omega t\right) $%
, $M=m_{1}+m_{2}$, $\mu =m_{1}m_{2}/M$, $\epsilon =m_{2}/M$. 

In the present work we put: 
\begin{equation}
R_{2}\left( y\right) =\alpha _{2}y-\gamma _{2}y^{3},\quad V_{2}\left( \dot{y}%
\right) =-\nu _{2}\dot{y},  \label{R2V2}
\end{equation}%
and assume that $\epsilon m_{2}$, $\nu _{1}$, $\alpha _{1}$ are small so
that the term proportional to $\epsilon m_{2}$ can be neglected.

Introducing nondimensional time $\tau $ and rescaling variable $y$: 
\begin{equation}
\tau =t\bar{\omega},\ \bar{\omega}=\sqrt{\frac{\alpha _{2}}{\mu }},\ z=y%
\sqrt{\frac{\gamma _{2}}{\alpha _{2}}},  \label{Ndim1}
\end{equation}%
we get the approximate effective equation \cite{Okninski2006}: 
\begin{equation}
\frac{d^{2}z}{d\tau ^{2}}+h\frac{dz}{d\tau }-z+z^{3}=-\gamma \frac{\Omega
^{2}}{\sqrt{\left( \Omega ^{2}-a\right) ^{2}+H^{2}\Omega ^{2}}}\cos \left(
\Omega \tau \right) ,  \label{eff}
\end{equation}%
with $\gamma \equiv G\frac{\kappa }{\kappa +1}$ and where nondimensional
constants are given by:%
\begin{equation}
h=\frac{\nu _{2}}{\mu \bar{\omega}},\ H=\frac{\nu _{1}}{M\bar{\omega}},\
\Omega =\frac{\omega }{\bar{\omega}},\ G=\frac{1}{\alpha _{2}}\sqrt{\frac{%
\gamma _{2}}{\alpha _{2}}}f,\ \kappa =\frac{m_{2}}{m_{1}},\ a=\frac{\alpha
_{1}\mu }{\alpha _{2}M}.  \label{Ndim2}
\end{equation}

We applied the Krylov-Bogoliubov-Mitropolsky (KBM) perturbation approach 
\cite{Nayfeh2011} to the effective equation (\ref{eff}) obtaining the
following implicit amplitude equation $F\left( A,\Omega \right) =0$ for the $1:1$ resonance \cite
{Okninski2006}:
\begin{equation}
A=\frac{\gamma \Omega ^{2}}{\sqrt{\left( h^{2}\Omega ^{2}+\left( 1+\Omega
^{2}-\frac{3}{4}A^{2}\right) ^{2}\right) \left( \left( \Omega ^{2}-a\right)
^{2}+H^{2}\Omega ^{2}\right) }}.  \label{Aeff}
\end{equation}

After a smooth change of variables,  $X\equiv \Omega ^{2}$, $Y\equiv A^{2}$, Eq. (\ref{Aeff}) 
can be written as:
\begin{equation}
L\left( X,Y\right) =\left( h^{2}X+\left( 1+X-\tfrac{3}{4}Y\right)
^{2}\right) \left( \left( X-a\right) ^{2}+H^{2}X\right) Y-\gamma ^{2}X^{2}=0.
\label{AEff}
\end{equation}

\section{Singular points of the amplitude profile and bifurcations of dynamics}
\label{General}
Let us consider a one degree of freedom dynamical system described by a
nonlinear, nonautonomous in general, ordinary differential equation of form:%
\begin{equation}
\frac{d^{2}y}{dt^{2}}+\omega ^{2}y=\varepsilon f\left( y,\frac{dy}{dt}%
,t\right) ,  \label{ODE}
\end{equation}%
where $\varepsilon $ is a small parameter and $f$ is a periodic function of
time with period $T=\frac{2\pi }{\Omega }$.

Applying the KBM method, or another asymptotic procedure, we can find an
approximate solution, the so-called $1:1$  nonlinear resonance, of form:
\begin{equation}
y\left( t\right) =A\cos \left( \Omega t+\varphi \right) +\varepsilon
y_{1}\left( t\right) +\ldots  \label{Solution}
\end{equation}%
where unknown amplitude $A$ and unknown frequency $\Omega $ fulfill the
amplitude equation:
\begin{equation}
F\left( A,\Omega ;\underline{c}\right) =0,  \label{AE}
\end{equation}
where $\underline{c}=\left( c_{1},c_{2},\ldots ,c_{n}\right) $ are
parameters. Equation (\ref{AE}) defines an implicit function -- a
two-dimensional planar curve - the amplitude profile. The form of this
curve, as well as stability of the solution (\ref{Solution}), determine
(approximately) dynamics of the system. In most cases the approximation (\ref%
{Solution}), (\ref{AE}) provides a good insight into dependence of the
dynamics on parameters.

It thus follows that qualitative changes of shape of the amplitude profile (%
\ref{AE}), referred henceforth as metamorphoses, induced by smooth changes
of control parameters $\underline{c}$, lead to qualitative changes of
dynamics (bifurcations) because dependence of the amplitude $A$ on frequency 
$\Omega $ changes qualitatively.

Stability of the solution (\ref{Solution}) is very important since it
determines which parts of the implicit curve (\ref{AE}) show up in actual
dynamics. Assume, for the sake of an example, that for fixed parameters $%
\underline{c}$ the curve $A\left( \Omega \right) $ has several branches in
the interval $\Omega \in \left( \Omega _{1},\Omega _{2}\right) $. Typically,
only some branches (or parts of these) are stable in this interval. 
The shape of the amplitude profile (\ref{AE}) (the stable part) can be
displayed by a bifurcation diagram, computed by numerical integration of the
equation (\ref{ODE}) in the interval $\Omega \in \left( \Omega _{1},\Omega
_{2}\right) $ for fixed parameters $\underline{c}$. Bifurcation diagrams,
computed for slowly changing parameters $\underline{c}$, can thus reveal
metamorphoses of the amplitude profiles.

According to the differential geometry of curves \cite{Spivak1965,Wall2004}\
an implicit curve changes its form at singular points which fulfill the
following equations: 
\begin{subequations}
\label{SINGULAR}
\begin{align}
F\left( A,\Omega ;\underline{c}\right) & =0,  \label{S1} \\
\frac{\partial F\left( A,\Omega ;\underline{c}\right) }{\partial A}& =0,
\label{S2} \\
\frac{\partial F\left( A,\Omega ;\underline{c}\right) }{\partial \Omega }&
=0.  \label{S3}
\end{align}
\end{subequations}
Solutions $A_{\ast }$, $\Omega _{\ast }$, $\underline{c}_{\ast }$ of Eqs. (%
\ref{SINGULAR}) are indeed special (singular). Equation (\ref{S2}) implies
that a function $A=g\left( \Omega \right) $ has no first derivative at $%
\Omega =\Omega _{\ast }$ while Eq. (\ref{S3}) means that a function $\Omega
=f\left( A\right) $ is not differentiable at $A=A_{\ast }$ \cite%
{Spivak1965,Wall2004}.

These conditions become more transparent when $F\left(
A,\Omega ;\underline{c}\right) $ is a polynomial in variables $A$, $\Omega $%
. In this case Eqs. (\ref{S1}), (\ref{S2}) mean that a polynomial $F\left(
A,\Omega _{\ast },\underline{c}_{\ast }\right) =a_{N}A^{N}+\ldots
+a_{1}A+a_{0}=0$ has multiple solutions (accordingly, a function $A=g\left(
\Omega \right) $ is not single-valued at $\Omega =\Omega _{\ast }$) while it
follows from equations (\ref{S1}), (\ref{S3}) that a polynomial $F\left(
A_{\ast },\Omega ,\underline{c}_{\ast }\right) =b_{M}\Omega ^{M}+\ldots
+b_{1}\Omega +b_{0}=0$ has multiple solutions (and a function $\Omega
=f\left( A\right) $ is not single-valued at $A=A_{\ast }$). Accordingly, $%
A_{\ast }$, $\Omega _{\ast }$, $\underline{c}_{\ast }$ is a multiple
solution of Eqs. (\ref{SINGULAR}).

Assume that a solution $\left( A_{\ast },\Omega _{\ast }\right) $ of Eqs. (%
\ref{SINGULAR}) exists for $\underline{c}=\underline{c}_{\ast }$ and there
are no other solutions in some neighbourhood of $\underline{c}_{\ast }$.
Consider a curve $\lambda $ in the parameter space containing the point $%
\underline{c}_{\ast }$. Then, when the curve $\lambda $ passes through the
point $\underline{c}_{\ast }$, the implicit function (\ref{AE}) changes its
shape. We shall refer to such a change as metamorphosis. This change depends
on the nature of a singular point. Two simplest examples of singular points
are isolated point and self-intersection. In an isolated point a new branch
of solution (stable or unstable) is born or disappears, while in a
self-intersection a branch (stable or unstable) becomes disconnected or two
branches are joined. In this work we compute the bifurcation set -- the set
in the parameter space containing all singular points of the amplitude
profile, see Ref. \cite{Kyziol2019} where the bifurcation set was computed
for the van der Pol-Duffing oscillator \cite{Kyziol2019}. 

The approach outlined in this Section can be generalized for $m:n$ resonances \cite{Kyziol2012} 
and for a system of coupled ordinary differential equations (\ref{gen-model}) \cite{Kyziol2015}. 
Investigation of changes of shape of
an amplitude profile induced by change of parameters was carried out in \cite%
{Holmes1976}\ for the Duffing equation. This is, however, a non-singular
case \cite{Kyziol2011} and thus only Eq. (\ref{S1}) and equivalent of Eq. (%
\ref{S2}) were used. The idea to use Implicit Function Theorem to
\textquotedblright define and find different branches intersecting at
singular points\textquotedblright\ of amplitude profiles was proposed in 
\cite{Awrejcewicz1995}.

\section{Global view of metamorphoses of the amplitude profiles}
\label{Global}
\subsection{Singular points}
\label{singular}

We shall investigate singular points of the amplitude equation (\ref{AEff})
because bifurcations occur at these points, cf. Section \ref{General}.
Singular points of algebraic curve $L\left( X,Y;a,h,H,\gamma \right) =0$ are
given by equations analogous to (\ref{SINGULAR}):
\begin{subequations}
\label{SING1}
\begin{eqnarray}
L &=&0,  \label{Sing1a} \\
\tfrac{\partial L}{\partial X} &=&0,  \label{Sing1b} \\
\tfrac{\partial L}{\partial Y} &=&0.  \label{Sing1c}
\end{eqnarray}
\end{subequations}
Equation (\ref{Sing1c}) factorizes as: 
\begin{equation}
\left. 
\begin{array}{l}
\frac{\partial L}{\partial Y}=\frac{1}{16}\left(
H^{2}X+X^{2}-2Xa+a^{2}\right)  \\ 
\times \left( 16X^{2}-48XY+16Xh^{2}+32X+27Y^{2}-48Y+16\right) =0. \label{factor}
\end{array}
\right. 
\end{equation}
It follows that the condition $H^{2}X+X^{2}-2Xa+a^{2}=0$ does not lead to
physical solutions \cite{Kyziol2011}, therefore instead of (\ref{Sing1c}) we
consider a simpler equation: 
\begin{equation}
M\left( X,Y;h\right) \equiv 16X^{2}-48XY+16Xh^{2}+32X+27Y^{2}-48Y+16=0.
\label{Sing1d}
\end{equation}%
Now we notice that the equation (\ref{Sing1d}) reduces to linear relations
between $X$ and $Y$ for two physical values of $h$: $h_{0}^{\left( 1\right)
}=\sqrt{\tfrac{4}{3}}$,$\ h_{0}^{\left( 2\right) }=0$. Indeed, we have: 
\begin{subequations}
\begin{eqnarray}
M\left( X,Y;h_{0}^{\left( 1\right) }\right)  &=&\tfrac{1}{3}\left(
4X-9Y+12\right) \left( 12X-9Y+4\right) =0,  \label{Sing1e} \\
M\left( X,Y;h_{0}^{\left( 2\right) }\right)  &=&\left( 4X-9Y+4\right) \left(
4X-3Y+4\right) =0\text{.}  \label{Sing1f}
\end{eqnarray}
\end{subequations}

\subsection{Bifurcation set}
\label{bifurcation}

It follows from general theory of implicit functions that in a singular
point there are multiple solutions of equation (\ref{AEff}). We shall use this
property to compute parameters values for which the amplitude profile 
defined by equation  (\ref{AEff}) has singular points. We shall refer to such 
set in the parameter space as the bifurcation set, see Ref. \cite{Holmes1976} 
where this term was used in the context of multiple solutions of the amplitude equation for the 
Duffing equation (albeit in the non-singular case).

To define a singular point we can use equations (\ref
{Sing1a}), (\ref{Sing1b}) and an alternative to condition (\ref{Sing1c}) 
which excludes existence of the single-valued function $Y=g\left( X\right)$. 
We thus solve Eqs. (\ref{Sing1a}), (\ref{Sing1b}) obtaining equation $%
f\left( X\right) =0$ for a function $Y=f\left( X\right) $ and then demand
that there are multiple solutions of this equation (alternatively, we could
have solved Eqs. (\ref{Sing1a}), (\ref{Sing1c})).

Solution of Eqs. (\ref{Sing1a}), (\ref{Sing1b}), with $L\left( X,Y;a,h,H,\gamma\right) $
given by Eq. (\ref{AEff}) can be written as the following equation for $X$: 
\begin{equation}
f\left( X\right)
=a_{8}X^{8}+a_{7}X^{7}+a_{6}X^{6}+a_{5}X^{5}+a_{4}X^{4}+a_{3}X^{3}+a_{2}X^{2}+a_{1}X+a_{0}=0,
\label{polynomial}
\end{equation}%
with coefficients $a_{0}$, $a_{1}$, \ldots , $a_{8}$ given in the \ref{A}.

Necessary and suficient condition for a polynomial to have multiple roots is
that its discriminant $\Delta $ vanishes \cite{Gelfand2008}, see also
lecture notes \cite{Janson2010}. Discriminant $\Delta $ can be computed as a
resultant of a polynomial $f\left( X\right) $ and its derivative $f^{\prime }
$, with a suitable normalizing factor.

Resultant  $R\left( f,g\right) $ of two polynomials, $f\left( X\right)
=a_{n}X^{n}+\ldots +a_{1}X+a_{0}$, $g\left( X\right) =b_{m}X^{m}+\ldots
+b_{1}X+b_{0}$, is given by determinant of the $\left( m+n\right) \times
\left( m+n\right) $ Sylvester matrix:%
\begin{equation}
R\left( f,g\right) =\det \left( 
\begin{array}{ccccccc}
a_{n} & a_{n-1} & a_{n-2} & \ldots  & 0 & 0 & 0 \\ 
0 & a_{n} & a_{n-1} & \ldots  & 0 & 0 & 0 \\ 
\vdots  & \vdots  & \vdots  &  & \vdots  & \vdots  & \vdots  \\ 
0 & 0 & 0 & \ldots  & a_{1} & a_{0} & 0 \\ 
0 & 0 & 0 & \ldots  & a_{2} & a_{1} & a_{0} \\ 
b_{m} & b_{m-1} & b_{m-2} & \ldots  & 0 & 0 & 0 \\ 
0 & b_{m} & b_{m-1} & \ldots  & 0 & 0 & 0 \\ 
\vdots  & \vdots  & \vdots  &  & \vdots  & \vdots  & \vdots  \\ 
0 & 0 & 0 & \ldots  & b_{1} & b_{0} & 0 \\ 
0 & 0 & 0 & \ldots  & b_{2} & b_{1} & b_{0}%
\end{array}%
\right) ,  \label{resultant}
\end{equation}%
see, for example, Eq. (1) in \cite{Janson2010}. Polynomials $f$ and $g$ have
a common root if and only if $R\left( f,g\right) =0$.

Therefore, the bifurcation set $\mathcal{M}$ is given by the following
equations:%
\begin{equation}
\left( a,h,H,\gamma \right) \in \mathcal{M}:\quad R\left( f,f^{\prime
}\right) =0,  \label{BifurcationSet}
\end{equation}%
where $R\left( f,f^{\prime }\right) $ is defined in (\ref{resultant}) 
as determinant of the $15 \times 15$ Sylvester matrix with 
$g=f^{\prime }$. The polynomial $f\left( X\right) $ has been defined in (\ref%
{polynomial}), with coefficients $a_{0}$, $a_{1}$, \ldots , $a_{8}$ given in
the \ref{A} and%
\begin{equation}
f^{\prime }\left( X\right) =8a_{8}X^{7}+7a_{7}X^{6}+\ldots
+2a_{2}X+a_{1}\equiv b_{7}X^{7}+b_{6}X^{6}+\ldots +b_{1}X+b_{0}.
\label{derivative}
\end{equation}%
Obviously, $R\left( f,f^{\prime }\right) $ depends on the parameters $
a,h,H,\gamma $ and will be also denoted as $R\left( f,f^{\prime }\right) \left( a,h,H,\gamma \right) $. 

\subsection{Plots of the bifurcation set}
\label{plots}

The equation $R\left( f,f^{\prime }\right) =0$ (\ref{BifurcationSet})  
leads to a high-order polynomial in variables $a,h,H,\gamma $. 
We were able, however, to plot $3D$ sections of the bifurcation set, 
$R\left( f,f^{\prime }\right) \left( a,h,H,\gamma \right) =0$, 
for fixed values of $h=h_0$ using the High Performance Computing Cluster (HPCC) at 
Politechnika \'{S}wi\c{e}tokrzyska.

The computations are much easier for $h_{0}^{\left( 1\right) }=\sqrt{\frac{4
}{3}}$ and $h_{0}^{\left( 2\right) }=0$ since then the polynomial (\ref
{polynomial}) factorizes into a product of two polynomials -- note that for
these values of $h$ the equation (\ref{Sing1d}) factorizes, see Eqs. (\ref
{Sing1e}), (\ref{Sing1f}). More exactly, for $h_{0}^{\left( 1\right) }=\sqrt{
\frac{4}{3}}$
\begin{eqnarray}
f\left( X\right) &=&f_{1}\left( X\right) f_{2}\left( X\right) ,
\label{factorization} \\
f_{1}\left( X\right) &=&b_{4}X^{4}+b_{3}X^{3}+b_{2}X^{2}+b_{1}X+b_{0}, 
\notag \\
f_{2}\left( X\right) &=&c_{4}X^{4}+c_{3}X^{3}+c_{2}X^{2}+c_{1}X+c_{0}, 
\notag
\end{eqnarray}%
where
\begin{eqnarray}
&&\left. 
\begin{array}{l}
b_{4}=144,\ b_{3}=-288a+144H^{2}+96 \\ 
b_{2}=144a^{2}-192a+96H^{2}-81\gamma ^{2}+16 \\ 
b_{1}=96a^{2}-32a+16H^{2},\ b_{0}=16a%
\end{array}%
\right\} ,  \label{f_1} \\
&&\left. 
\begin{array}{l}
c_{4}=16,\ c_{3}=-32a+16H^{2}+96 \\ 
c_{2}=16a^{2}-192a+96H^{2}+144 \\ 
c_{1}=96a^{2}-288a+144H^{2}-81\gamma ^{2},\ c_{0}=144a^{2}%
\end{array}%
\right\} ,  \label{f_2}
\end{eqnarray}%
while for $h_{0}^{\left( 2\right) }=0$ we have:
\begin{eqnarray}
f\left( X\right) &=&-3\gamma ^{2}\,X\,g\left( X\right) ,  \label{g} \\
g\left( X\right) &=&d_{5}X^{5}+d_{4}X^{4}+d_{3}X^{3}+d_{2}X^{2}+d_{1}X+d_{0},
\notag
\end{eqnarray}%
where%
\begin{equation}
\left. 
\begin{array}{l}
d_{5}=16,\ d_{4}=16H^{2}-32a+48 \\ 
d_{3}=48+16a^{2}-96a+48H^{2} \\ 
d_{2}=48H^{2}+48a^{2}-81\gamma ^{2}+16-96a \\ 
d_{1}=-32a+48a^{2}+16H^{2},\ d_{0}=16a^{2}
\end{array}
\right\} .  \label{quintic}
\end{equation}

\begin{figure}[h!]
\center 
\includegraphics[width=12cm, height=8cm]{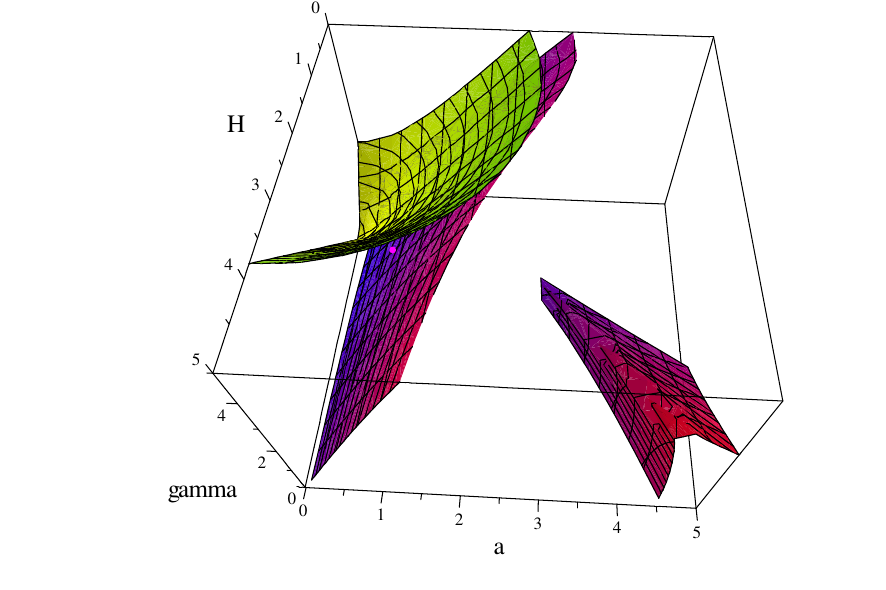}
\includegraphics[width=12cm, height=8cm]{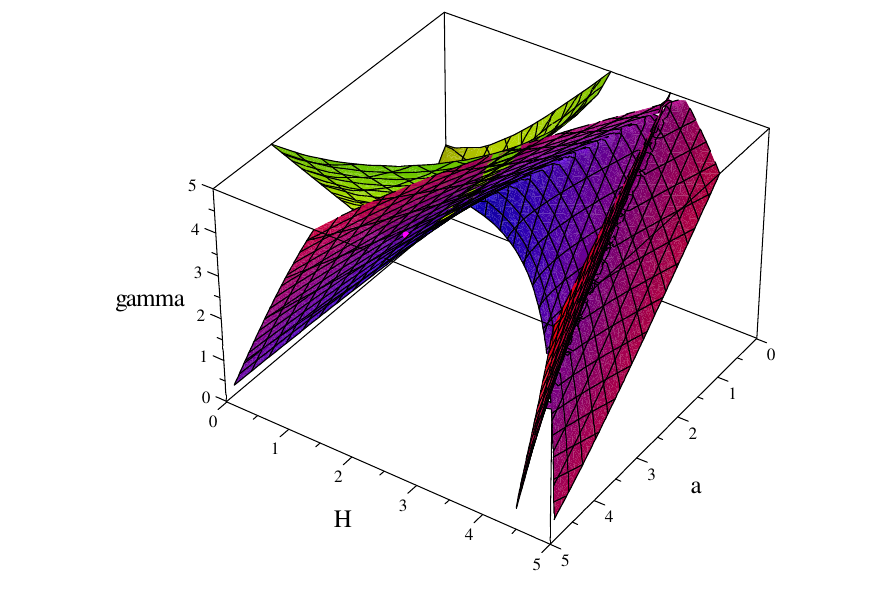}
\caption{The bifurcation subsets $\left( h_{0}^{(1)}=\sqrt{\frac{4}{3}}\right) $:$%
\quad \Delta _{1}\left( a,H,\gamma \right) =0$ blue-red and $R\left(
f_{1},f_{2}\right) =0$ yellow-green (top) and $\Delta _{2}\left( a,H,\gamma
\right) =0$ blue-red and $R\left( f_{1},f_{2}\right) =0$ yellow-green
(bottom). Magenta dots mark singular points described in Section \ref{Numerical}.}
\label{FF2}
\end{figure}

Discriminant of a quartic polynomial $f\left( x\right)
=ax^{4}+bx^{3}+cx^{2}+dx+e$ is:
\begin{eqnarray}
\Delta
&=&256a^{3}e^{3}-192a^{2}bde^{2}-128a^{2}c^{2}e^{2}+144a^{2}cd^{2}e-27a^{2}d^{4}
\label{quartic} \\
&&+144ab^{2}ce^{2}-6ab^{2}d^{2}e-80abc^{2}de+18abcd^{3}+16ac^{4}e  \notag \\
&&-4ac^{3}d^{2}-27b^{4}e^{2}+18b^{3}cde-4b^{3}d^{3}-4b^{2}c^{3}e+b^{2}c^{2}d^{2}
\notag
\end{eqnarray}
Therefore, the bifurcation set for $h_{0}^{\left( 1\right) }=0$ consists of three subsets: $%
\Delta _{1}\left( a,H,\gamma \right) =0$, $\Delta _{2}\left( a,H,\gamma
\right) =0$ where $\Delta _{1}$, $\Delta _{2}$ are discriminants of
polynomials $f_{1}$, $f_{2}$, respectively, and $R\left( f_{1},f_{2}\right)
=0$ where the resultant $R\left( f_{1},f_{2}\right) $ reads:
\begin{equation}
\left. 
\begin{array}{l}
R\left( f_{1},f_{2}\right) =\det \left( 
\begin{array}{cccccccc}
b_{4} & b_{3} & b_{2} & b_{1} & b_{0} & 0 & 0 & 0 \\ 
0 & b_{4} & b_{3} & b_{2} & b_{1} & b_{0} & 0 & 0 \\ 
0 & 0 & b_{4} & b_{3} & b_{2} & b_{1} & b_{0} & 0 \\ 
0 & 0 & 0 & b_{4} & b_{3} & b_{2} & b_{1} & b_{0} \\ 
c_{4} & c_{3} & c_{2} & c_{1} & c_{0} & 0 & 0 & 0 \\ 
0 & c_{4} & c_{3} & c_{2} & c_{1} & c_{0} & 0 & 0 \\ 
0 & 0 & c_{4} & c_{3} & c_{2} & c_{1} & c_{0} & 0 \\ 
0 & 0 & 0 & c_{4} & c_{3} & c_{2} & c_{1} & c_{0}%
\end{array}%
\right) \medskip  \\ 
=2^{12}3^{8}\gamma ^{4}a^{2}\left( 256a^{2}-512a+256H^{2}-81\gamma
^{2}+256\right) ^{3}%
\end{array}%
\right.   \label{res}
\end{equation}
These three subsets are plotted in Fig. (\ref{FF2}). 
Discriminant of the quintic polynomial is more complicated: 
\begin{equation}
\Delta =a_{5}^{-1}\det \left( 
\begin{array}{ccccccccc}
a_{5} & a_{4} & a_{3} & a_{2} & a_{1} & a_{0} & 0 & 0 & 0 \\ 
0 & a_{5} & a_{4} & a_{3} & a_{2} & a_{1} & a_{0} & 0 & 0 \\ 
0 & 0 & a_{5} & a_{4} & a_{3} & a_{2} & a_{1} & a_{0} & 0 \\ 
0 & 0 & 0 & a_{5} & a_{4} & a_{3} & a_{2} & a_{1} & a_{0} \\ 
5a_{5} & 4a_{4} & 3a_{3} & 2a_{2} & a_{1} & 0 & 0 & 0 & 0 \\ 
0 & 5a_{5} & 4a_{4} & 3a_{3} & 2a_{2} & a_{1} & 0 & 0 & 0 \\ 
0 & 0 & 5a_{5} & 4a_{4} & 3a_{3} & 2a_{2} & a_{1} & 0 & 0 \\ 
0 & 0 & 0 & 5a_{5} & 4a_{4} & 3a_{3} & 2a_{2} & a_{1} & 0 \\ 
0 & 0 & 0 & 0 & 5a_{5} & 4a_{4} & 3a_{3} & 2a_{24} & a_{1}%
\end{array}%
\right) ,  \label{Delta3}
\end{equation}
see Eq. (1.36) in Chapter 12 in \cite{Gelfand2008} for explicit formula. 

\begin{figure}[h!]
\center 
\includegraphics[width=12cm, height=8cm]{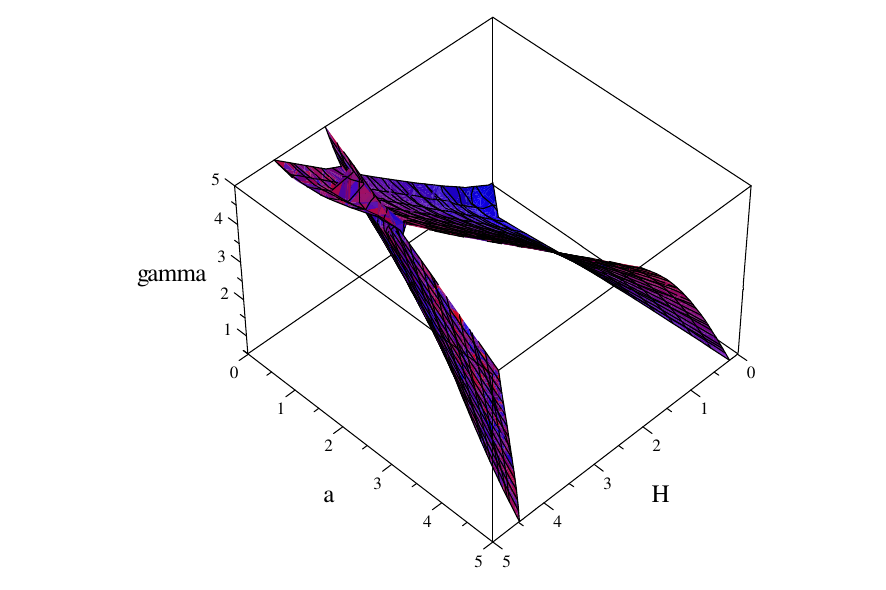}
\caption{The bifurcation set: $\Delta\left( a,H,\gamma\right) =0$, $h^{(2)}_0=0$.}
\label{FF3}
\end{figure}

The bifurcation set $\Delta\left( a,H,\gamma \right) =0$, where $\Delta$ 
is discriminant (\ref{Delta3}) of polynomial $g\left( X\right)$, see Eq. (\ref{g}),  is plotted in Fig. \ref{FF3}.

Moreover, for $h_{0}^{\left( 1\right) }=\sqrt{\frac{4}{3}}$ we were able to compute degenerate
singular points, i.e. fulfilling Eqs. (\ref{SING1}) and additional
condition:
\begin{equation}
\frac{\partial ^{2}L}{\partial X^{2}}\frac{\partial ^{2}L}{\partial Y^{2}}%
-\left( \frac{\partial ^{2}L}{\partial X\partial Y}\right) ^{2}=0,
\label{Degenerate1}
\end{equation}
which implies that determinant of the Hesse matrix vanishes.The physical solution reads: 
\begin{equation}
\left. 
\begin{array}{l}
h_{\ast }=h_{0}^{\left( 1\right) }=\sqrt{\frac{4}{3}},\ H_{\ast}=\frac{1}{32}\sqrt{
-2048+162\gamma ^{2}+64\sqrt{1024+162\gamma ^{2}}}, \\ 
a_{\ast }=-1+\sqrt{4+H^{2}},\ \gamma \text{ -- arbitrary},\ X_{\ast }=1,\
Y_{\ast }=\frac{16}{9}.%
\end{array}%
\right\}   \label{Degenerate2}
\end{equation}
Let us note that these parameters $a_{\ast }$, $H_{\ast }$, $\gamma $ are
also a solution of equations:
\begin{equation}
\Delta _{1}\left( a,H,\gamma \right) =0,\quad \Delta _{2}\left( a,H,\gamma
\right) =0,\quad R\left( f_{1},f_{2}\right) =0,  \label{Degenerate3}
\end{equation}
hence three surfaces in Fig. (\ref{FF2}) are tangent along line 
$\left( a_{\ast },H_{\ast },\gamma \right) $. 

\begin{figure}[h!]
\center 
\includegraphics[width=12cm, height=8cm]{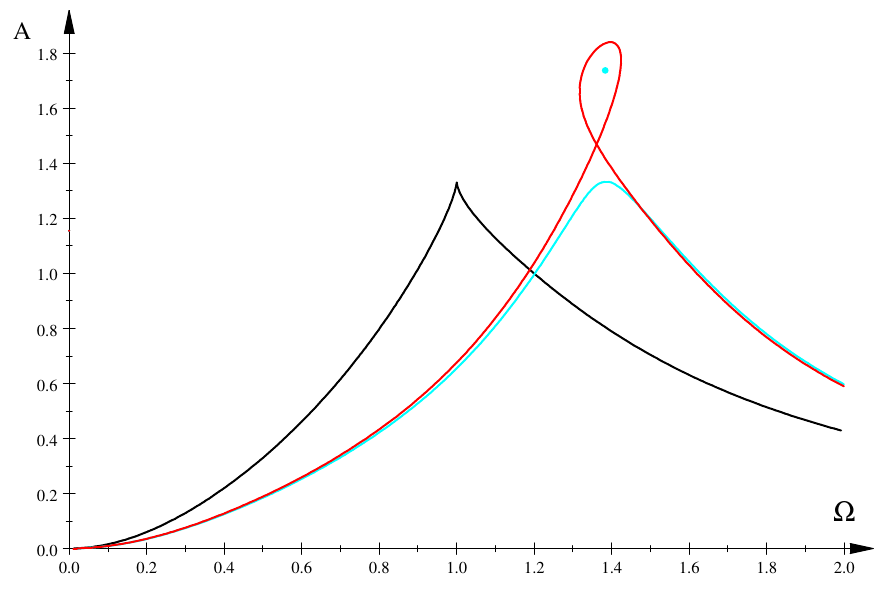}
\caption{The amplitude profiles: $a=a_{\ast }$, $h=h_{\ast }$, $H=H_{\ast }$, $\gamma
=\gamma _{\ast }$ (black, degenerate singular), $h=\sqrt{\frac{4}{3}}$, $%
H=0.9$, $\gamma =2$, $a_{1}=2.199\,923$ (blue-green, isolated point), $%
a_{2}=2.156\,818$ (red, self-intersection).}
\label{FF4}
\end{figure}

We shall now determine the form of the amplitude profile $F\left( A,\Omega
\right) =0$ in the case of a degenerate singular point and for two singular
points in its vicinity in the parameter space. Let $h_{\ast }=\sqrt{\frac{4}{
3}}$. Then, we have from Eq. (\ref{Degenerate2}) $a_{\ast }=1.\,193\,171$, $
H_{\ast}=0.9$, $\gamma_{\ast} =1.\,636\,439$. The amplitude profile in the case of
degenerate singular point has a cusp, see black line in Fig. \ref{FF4}.

There are also two singular non-degenerate points shown in Fig. \ref{FF4}: self intersection (red
line) and isolated point (blue-green). 
The parameters were chosen as $h=\sqrt{\frac{4}{3}}$, $H=0.9$, $\gamma =2$.
Solving equation $\Delta _{1}=0$ we get $a=2.\,199\,923$ -- it is an
isolated point, while $\Delta _{2}=0$ yields $a=2.\,156\,818$ -- a
self-intersection, see also magenta dots in Fig.  \ref{FF2}.

\begin{figure}[h!]
\center 
\includegraphics[width=10cm, height=10cm]{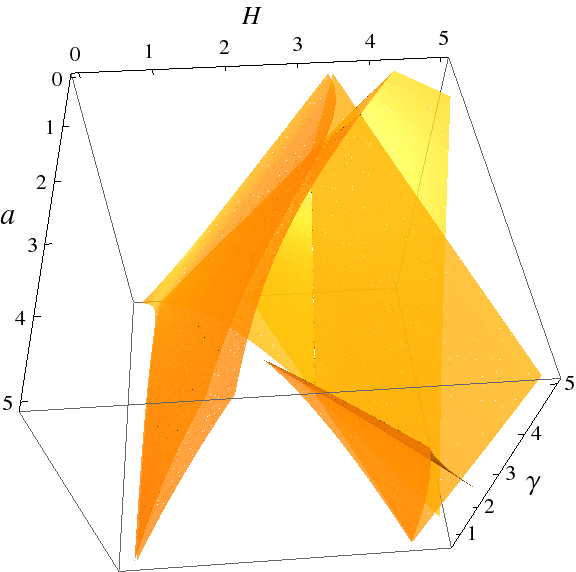}
\caption{The bifurcation set, $h^{(3)}_0 = 1$.}
\label{FF5}
\end{figure}

Sections of the bifurcation set displayed in Figs. \ref{FF2}, \ref{FF3} 
has been computed for $h^{(1)}_0=\sqrt{\frac{4}{3}}$, $h^{(2)}_0=0$, respectively,  since then 
the polynomial (\ref{polynomial}) factorizes and computations are simpler. For the sake of comparison, 
we have also computed and plotted the bifurcation set for $h^{(3)}_0=1$, using the HPCC, see Fig. \ref{FF5}.

\newpage

\section{Examples of bifurcations at singular points of the amplitude profile}
\label{Numerical}
We shall demonstrate that knowledge of singular points of the amplitude profiles $F\left( A,\Omega \right) =0$, 
computed by application of the KBM method and defined implicitly in  Eq. (\ref{Aeff}) or (\ref{AEff}), 
permits prediction of metamorphoses of dynamics. 

Global views of singular points are displayed in Figs. \ref{FF2}, \ref{FF3}, \ref{FF5}, 
showing $\mathcal{M}\left( a,h_{0},H,\gamma \right)$ -- sections of the bifurcation set defined in 
Eq. (\ref{BifurcationSet}). More exactly, for each point $\left( a,h_{0},H,\gamma \right) \in \mathcal{M}
$ the amplitude profile $F\left( A,\Omega ;a,h_{0},H,\gamma \right) =0$ has a
singular point.

Consider, for example, the case  $h^{(1)}_0=\sqrt{\frac{4}{3}}$. It follows from Figs. (\ref{FF2}), (\ref{FF4}) 
that in the  neighbourhood of a degenerate singular
points $\left( a_{\ast },h_{\ast },H_{\ast},\gamma\right)$ there are two
kinds of singular points:  isolated points and self-intersections. In the
neighbourhood of the first singular point 
$\left( a,h,H,\gamma \right) =\left( 2.\,156\,818,h_{0}^{\left( 1\right)},0.9,2\right) $
the amplitude profiles are shown in Fig. \ref{FF6}:

\begin{figure}[h!]
\center 
\includegraphics[width=6cm, height=4cm]{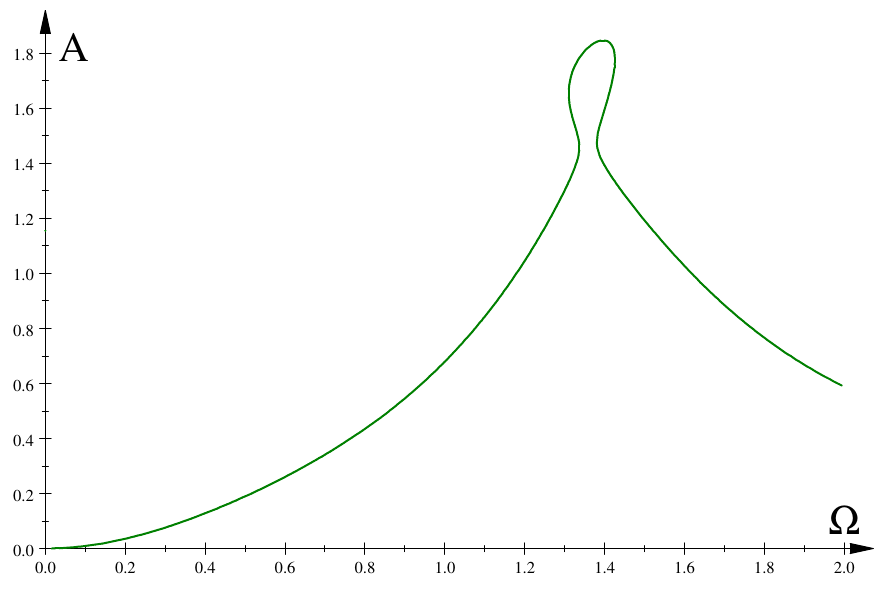}
\includegraphics[width=6cm, height=4cm]{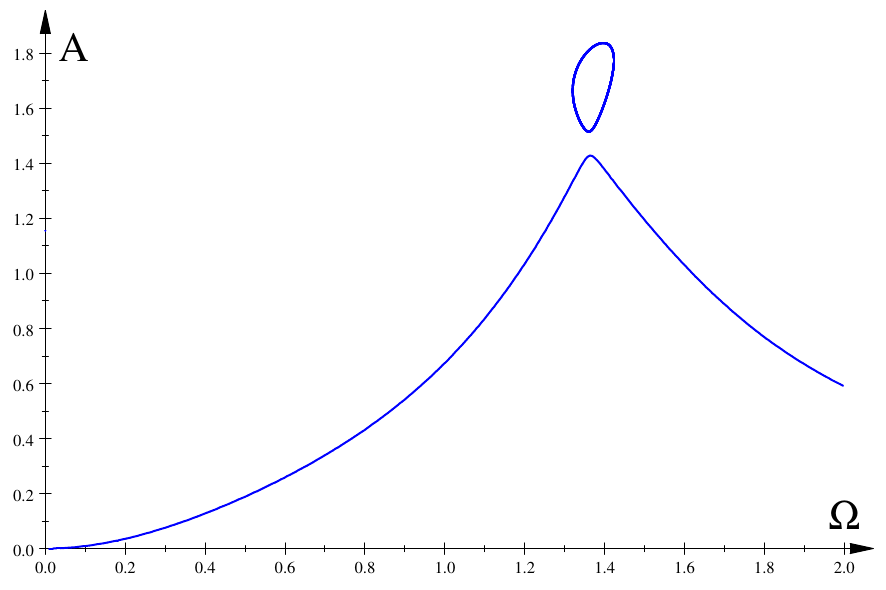}
\caption{The amplitude profiles: $h=h_{0}^{\left( 1\right) }$, $H=0.9$, $\gamma =2$, $%
a=2.\,15$ (left, green), $a=2.16$ (right, blue).}
\label{FF6}
\end{figure}

Bifurcation diagrams \cite{Nusse1997} show,  indeed, metamorphosis of dynamics, see Fig. \ref{FF7}:
\begin{figure}[h!]
\center
\includegraphics[width=6cm, height=4cm]{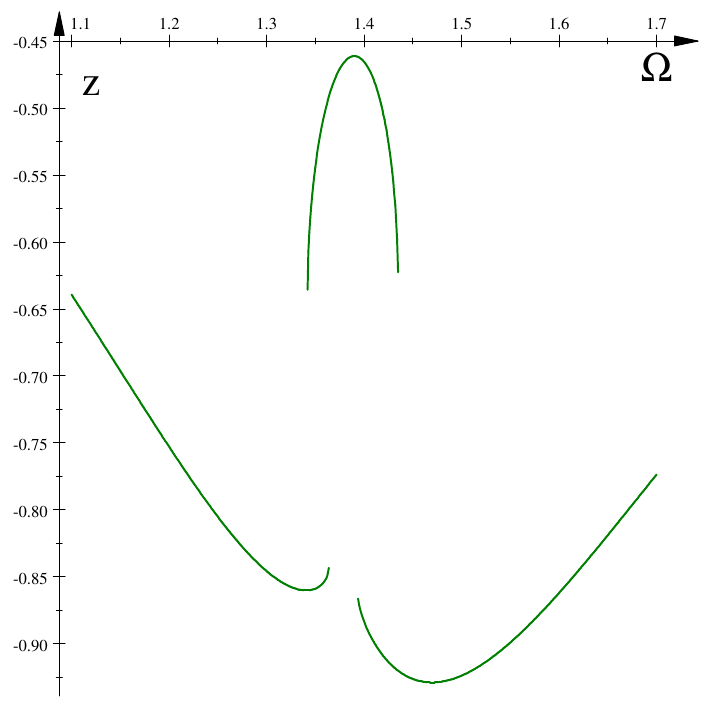}
\includegraphics[width=6cm, height=4cm]{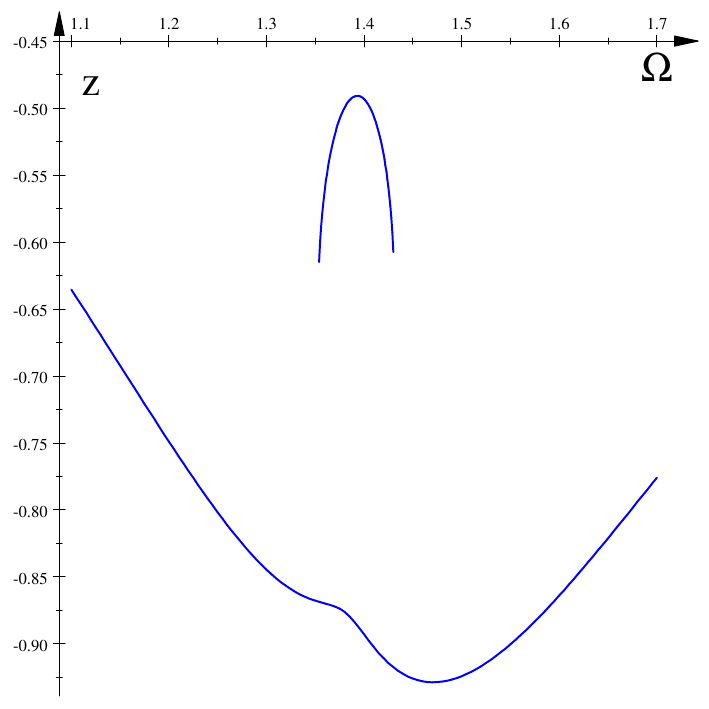}
\caption{The bifurcation diagrams: $h=h_{0}^{\left( 1\right) }$, $H=0.9$, $\gamma =2$, $%
a=2.\,21$ (left, green), $a=2.22$ (right, blue).}
\label{FF7}
\end{figure}

Fig. \ref{FF7} shows clearly which parts of the amplitude profiles $F\left( A,\Omega \right) =0$ 
shown in Fig. \ref{FF6} are stable. Please note that in the computed bifurcation diagrams only 
stable solutions are displayed.

\begin{figure}[h!]
\center 
\includegraphics[width=6cm, height=4cm]{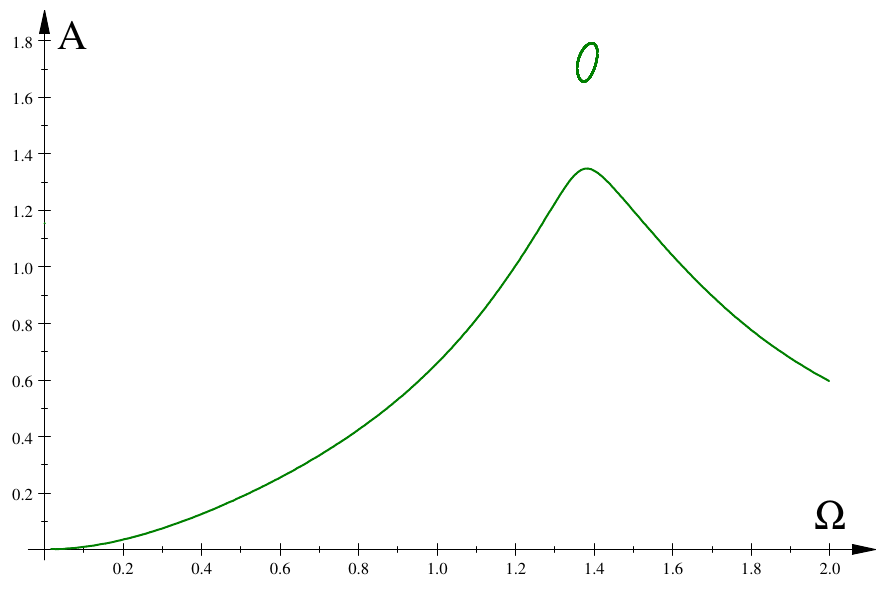}
\includegraphics[width=6cm, height=4cm]{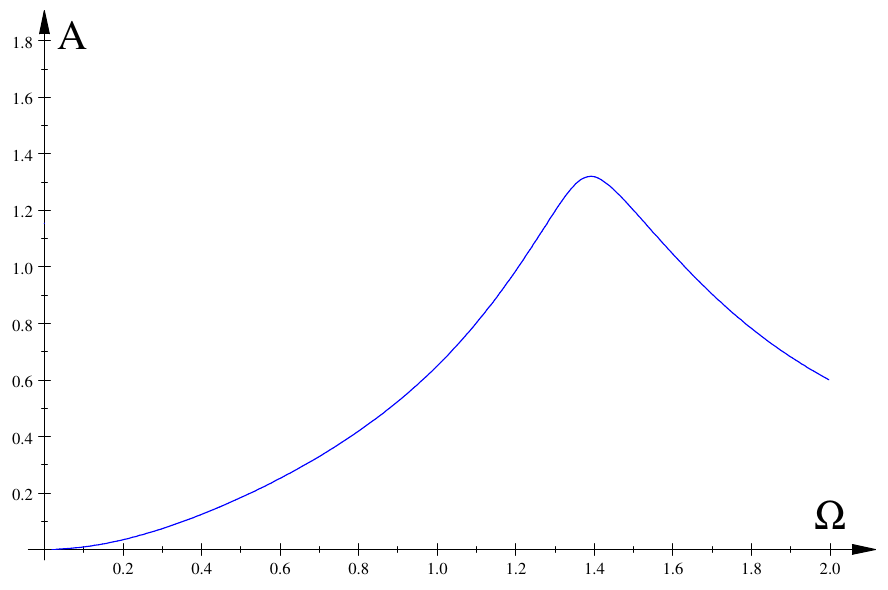}
\caption{The amplitude profiles: $h=h_{0}^{\left( 1\right) }$, $H=0.9$, $\gamma =2$, $%
a=2.\,19$ (left, green), $a=2.21$ (right, blue).}
\label{FF8}
\end{figure}

\begin{figure}[h!]
\center
\includegraphics[width=6cm, height=4cm]{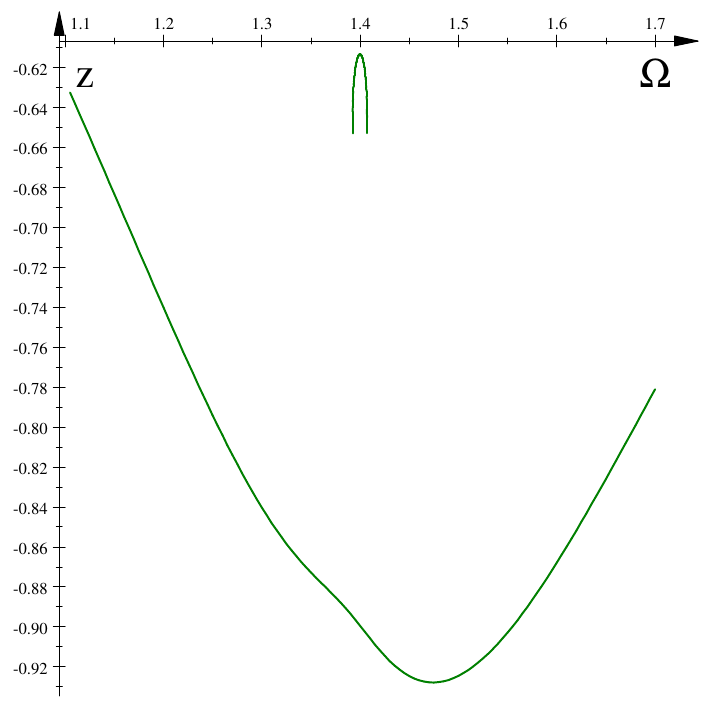}
\includegraphics[width=6cm, height=4cm]{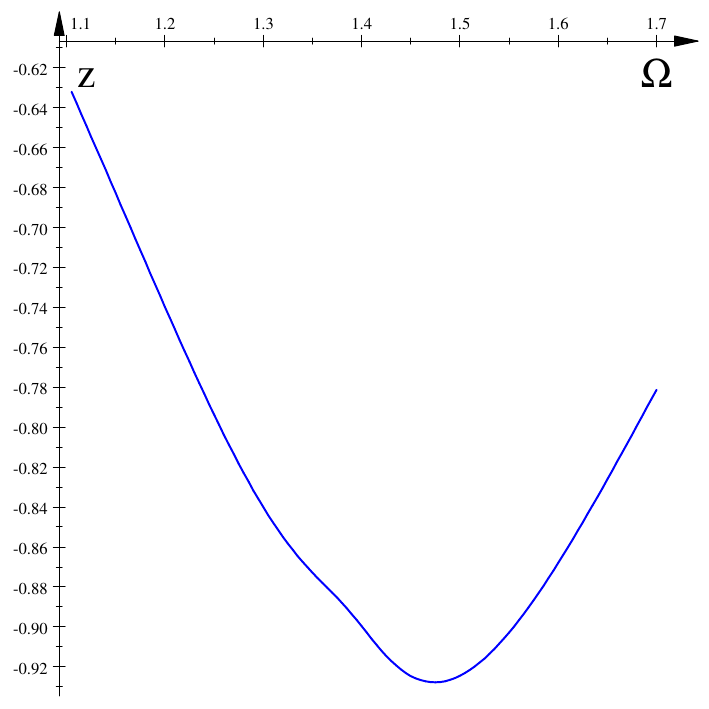}
\caption{The bifurcation diagrams:  $h=h_{0}^{\left( 1\right) }$, $H=0.9$, $\gamma =2$, $%
a=2.\,242$ (left, green), $a=2.243$ (right, blue).}
\label{FF9}
\end{figure}

Near the second singular point $\left( a,h,H,\gamma
\right) =\left( 2.\,199\,923,\ h^{(1)}_0,\ 0.9,\ 2\right) $ the amplitude profiles 
are shown in Fig. \ref{FF8} and the corresponding bifurcation diagrams are displayed in Fig. \ref{FF9}. 
Figs. \ref{FF8}, \ref{FF9} show metamorphosis of dynamics -- disappearance of 
a branch of the solution.

\section{Summary and discussion}
\label{Summary}

In this work we have studied metamorphoses of amplitude profiles for the
effective equation (\ref{eff}) which describes approximately dynamics of two coupled
periodically driven oscillators. Our computations are based on the amplitude profile 
$F\left( A,\Omega \right) =0$ computed  in \cite{Kyziol2011} within the approximate KBM
method. 
We have demonstrated in our earlier papers that the
amplitude profiles $F\left( A,\Omega \right) =0$ computed by the KBM method 
permit prediction of metamorphoses of dynamics which occur at singular points. 

In the present study we have strived at a global view of singular points of
the amplitude profiles. We have derived equations for the bifurcation set,
see Eqs. (\ref{BifurcationSet}), (\ref{resultant}), (\ref{polynomial}) and the \ref{A}. The bifurcation
set is thus a four-dimensional manifold $\mathcal{M}$ containing all points $%
\left( a,h,H,\gamma \right) $ for which the amplitude profile $F\left(
A,\Omega ;a,h,H,\gamma \right) =0$ is singular. Examples of plots of $%
\mathcal{M}$ and the corresponding bifurcations are given in Sections \ref{plots} and \ref{Numerical},
respectively.

\appendix{} 
\section{Coefficients of the polynomial (\ref{polynomial})}
\label{A}

$a_{0}=64a^{4}h^{2}$

$a_{1}= 
128h^{4}a^{4}+128H^{2}h^{2}a^{2}-48a^{2}
\gamma^{2}-256a^{3}h^{2}+256a^{4}h^{2}$

$a_{2}=\left( 
\begin{array}{l}
64h^{6}a^{4}+384a^{4}h^{2}+256h^{4}a^{4}+96\gamma
^{2}a+64H^{4}h^{2}+384h^{2}a^{2} \\ 
-144a^{2}\gamma
^{2}-512h^{4}a^{3}-256H^{2}h^{2}a+256h^{4}H^{2}a^{2}+512H^{2}h^{2}a^{2} \\ 
-432\gamma ^{2}h^{2}a^{2}-48\gamma ^{2}H^{2}-1024a^{3}h^{2}%
\end{array}%
\right) $

$a_{3}=\left( 
\begin{array}{l}
128h^{4}a^{4}-256h^{2}a+128h^{2}H^{2}-256h^{6}a^{3}-1024h^{4}a^{3} \\ 
+256H^{4}h^{2}-432\gamma ^{2}h^{2}a^{2}-144\gamma
^{2}H^{2}+768H^{2}h^{2}a^{2}+1536h^{2}a^{2} \\ 
-144a^{2}\gamma ^{2}+864\gamma
^{2}h^{2}a-1536a^{3}h^{2}+768h^{4}a^{2}+243\gamma ^{4} \\ 
+256a^{4}h^{2}-48\gamma ^{2}+288\gamma ^{2}a-512h^{4}aH^{2}-432\gamma
^{2}h^{2}H^{2} \\ 
-1024H^{2}h^{2}a+128h^{4}H^{4}+512h^{4}H^{2}a^{2}+128h^{6}H^{2}a^{2}%
\end{array}%
\right) $

$a_{4}=\left( 
\begin{array}{l}
64h^{2}+288\gamma ^{2}a-144\gamma
^{2}H^{2}+2304h^{2}a^{2}-1024h^{2}a+512h^{2}H^{2} \\ 
-432\gamma
^{2}h^{2}+64a^{4}h^{2}+384h^{6}a^{2}-512h^{4}a+1536h^{4}a^{2}-144\gamma ^{2}
\\ 
-256h^{6}aH^{2}+864\gamma ^{2}h^{2}a+512H^{2}h^{2}a^{2}-432\gamma
^{2}h^{2}H^{2}-1536H^{2}h^{2}a \\ 
-1024h^{4}aH^{2}+256h^{4}H^{2}a^{2}+64h^{6}H^{4}-512h^{4}a^{3}+256h^{4}H^{2}
\\ 
+256h^{4}H^{4}-1024a^{3}h^{2}+384H^{4}h^{2}-48a^{2}\gamma ^{2}%
\end{array}%
\right) $

$a_{5}=\left( 
\begin{array}{l}
256H^{4}h^{2}-48\gamma
^{2}H^{2}+128h^{4}-256a^{3}h^{2}+512h^{4}H^{2}+1536h^{2}a^{2} \\ 
+768h^{4}a^{2}-1536h^{2}a-432\gamma ^{2}h^{2}+96\gamma
^{2}a+256h^{2}-1024h^{4}a \\ 
-144\gamma
^{2}-256h^{6}a-512h^{4}aH^{2}-1024H^{2}h^{2}a+128h^{6}H^{2}+128h^{4}H^{4} \\ 
+768h^{2}H^{2}+128H^{2}h^{2}a^{2}%
\end{array}%
\right) $

$a_{6}=\left( 
\begin{array}{l}
512h^{2}H^{2}+64h^{6}+384h^{2}a^{2}-48\gamma ^{2}-512h^{4}a+384h^{2} \\ 
+64H^{4}h^{2}-1024h^{2}a-256H^{2}h^{2}a+256h^{4}+256h^{4}H^{2}%
\end{array}%
\right) $

$a_{7}=128h^{4}-256h^{2}a+256h^{2}+128h^{2}H^{2}$

$a_{8}=64h^{2}$

\section{Computational details}
\label{Details}
Nonlinear polynomial equations were solved numerically using the computational engine Maple 4.0 
from the Scientific WorkPlace 4.0. 
Fig. \ref{FF5} was plotted by the computational engine Mathematica 9.0 installed in the High Performance 
Computing Cluster at Politechnika \'{S}wi\c{e}tokrzyska (Kielce University of Technology). 
Other figures were plotted with the computational engine MuPAD 4.0 from Scientific WorkPlace 5.5. 
Curves shown in bifurcation diagrams in Figs. \ref{FF7}, \ref{FF9} were computed  by 
integrating numerically Eq. (\ref{eff}) running DYNAMICS, 
program written by Helena E. Nusse and James A. Yorke \cite{Nusse1997}, and our own programs written in Pascal.


\begin{thebibliography}{99}

\bibitem{Kyziol2017} J. Kyzio{\l }, A. Okni\'{n}ski, Metamorphoses of
resonance curves in systems of coupled oscillators: The case of degenerate
singular points, \textit{Int. J. Non-Linear Mechanics} \textbf{95}, 272--276 (2017).

\bibitem{Mahmoud2004} G.M. Mahmoud, T. Bountis, The dynamics of systems of complex nonlinear oscillators: a review, 
\textit{Int. J. Bifur. Chaos} \textbf{14} (2004) 3821-3846.

\bibitem{Pikovsky2015} A. Pikovsky, M. Rosenblum, Dynamics of globally coupled oscillators: 
Progress and perspectives, \textit{Chaos}, \textbf{25} (2015) 097616.

\bibitem{Schultheiss2011} N.W. Schultheiss, A.A. Prinz, R.J. Butera, eds.
\textit{Phase response curves in neuroscience: theory, experiment, and analysis}. Springer 
A Science \& Business Media, 2011. 

\bibitem{Awal2019} N.M. Awal, D. Bullara, I.R. Epstein, The smallest chimera, Periodicity and chaos 
in a pair of coupled chemical oscillators, \textit{Chaos} \textbf{29} (2019) 013131. 

\bibitem{Hajjaj2019} A.Z. Hajjaj, N. Jaber, S. Ilyas, F.K. Alfosail, M.I. Younis, 
Linear and nonlinear dynamics of micro- and nano-resonators: Review of recent advances, 
\textit{Int. J. Non-Linear Mechanics}, 2019, March 2020, 103328, https://doi.org/10.1016/j.ijnonlinmec.2019.103328.

\bibitem{Kozlowski1995} J. Koz{\l }owski, U. Parlitz and W. Lauterborn, 
Bifurcation analysis of two coupled periodically driven Duffing oscillators, 
\textit{Phys. Rev.} E \textbf{51} (1995) 1861-1867.

\bibitem{Kuznetsov2009} A. P. Kuznetsov, N. V. Stankevich and L. V Turukina, 
Coupled ven der Pol-Duffing oscillators: phase dynamics and structure of 
synchronization tongues, 
\textit{Physica} D \textbf{238} (2009)  1203-1215.

\bibitem{Perkins2012} E. Perkins, B. Balachandran, Noise-enhanced response of nonlinear oscillators, 
\textit{Procedia Iutam} \textbf{5} (2012) 59-68.

\bibitem{Sabrathinam2013} S. Sabarathinam, K. Thamilmaran, L. Borkowski, P. Perlikowski, 
P. Brzeski, A. Stefanski, T. Kapitaniak, Transient chaos in two coupled, dissipatively perturbed Hamiltonian Duffing 
oscillators, \textit{Commun. Nonlinear Sci Numer. Simulat.} \textbf{18} (2013) 3098-3107.

\bibitem{Zulli2016} D. Zulli, A. Luongo, Control of primary and subharmonic resonances of a Duffing oscillator 
via non-linear energy sink, \textit{Int. J. Non-Linear Mechanics} \textbf{80} (2016) 170-182.

\bibitem{Luo2017} B. Yu, A.C.J. Luo, Analytical period-1 motions to chaos in a two-degree-of-freedom
oscillator with a hardening nonlinear spring, \textit{Int. J. Dynam. Control} \textbf{5} (2017) 436–453.

\bibitem{Karahan2017} M.M. Faith Karahan, M. Pakdemirli, Free and forced vibrations of the strongly nonlinear 
cubic-quintic Duffing oscillators, \textit{Zeit. Natur.} A \textbf{72} (2017) 59-69.

\bibitem{Papangelo2019} A. Papangelo, F. Fontanela, A. Grolet, M. Ciavarella, N. Hoffmann, Multistability and 
localization in forced cyclis structures modelled by weakly-coupled Duffing oscillators, 
\textit{J. Sound Vibr.} \textbf{440} (2019) 202-211.

\bibitem{DenHartog1985} J. P. Den Hartog, \textit{Mechanical Vibrations}
(4th edition), Dover Publications, New York 1985.

\bibitem{Oueini1999} S. S. Oueini, A. H. Nayfeh and J.R. Pratt, A review of development and implementation of 
an active nonlinear vibration absorber\textit{Arch. Appl. Mech.}, \textbf{69} (1999) 585-620.

\bibitem{Nayfeh2011} A.H.~Nayfeh, \textit{Introduction to Perturbation Techniques}, John Wiley \& Sons, 2011.

\bibitem{Okninski2006} A.~Okni\'{n}ski and J.~Kyzio\l , Perturbation
analysis of the effective equation for two coupled periodically driven
oscillators, \textit{Diff. Eqs. Nonlin. Mec.} \textbf{2006} (2006) 56146.

\bibitem{Kyziol2019} J.~Kyzio\l\ and A.~Okni\'{n}ski, Van der Pol-Duffing
oscillator: Global view of metamorphoses of the amplitude profiles, 
\textit{Int. J.Nonlinear Mech.} \textbf{116} (2019) 102-106.

\bibitem{Kyziol2013b} J.~Kyzio\l\ and A.~Okni\'{n}ski, Exact nonlinear
fourth-order equation for two coupled oscillators: metamorphoses of
resonance curves, \textit{Acta Phys. Polon.} B \textbf{44} (2013) 35-47.

\bibitem{Spivak1965} M. Spivak, \textit{Calculus on Manifolds}, W.A.
Benjamin, Inc., Menlo Park (California) 1965.

\bibitem{Wall2004} C. T. C. Wall, \textit{Singular Points of Plane Curves},
Cambridge University Press, New York 2004.

\bibitem{Kyziol2012} J.~Kyzio\l\ and A.~Okni\'{n}ski, Coupled nonlinear oscillators: 
metamorphoses of amplitude profiles for the approximate effective equation -- 
the case of $1:3$ resonance, \textit{Acta Phys. Polon.} B 43 (2012) 1275-1287. 

\bibitem{Kyziol2015} J.~Kyzio\l , Metamorphoses of resonance curves for two
coupled oscillators: The case of small non-linearities in the main mass
frame, Int. J. Nonlinear Mech. 76 (2015) 164--168.

\bibitem{Holmes1976} P.J. Holmes, D.A. Rand, The bifurcations of Duffing's
equation: an application of Catastrophe Theory, Journal of Sound and
Vibration 44 (1976) 237-253.

\bibitem{Kyziol2011} J.~Kyzio\l\ and A.~Okni\'{n}ski, Coupled nonlinear
oscillators: metamorphoses of amplitude profiles. The case of the
approximate effective equation, \textit{Acta Phys. Polon.}  B \textbf{42} (2011) 2063-2076.

\bibitem{Awrejcewicz1995} J. Awrejcewicz, Modified Poincar\'{e} method and
implicit function theory, in: \textit{Nonlinear Dynamics: New Theoretical
and Applied Results}, J. Awrejcewicz (ed.), Akademie Verlag, Berlin, 1995,
pp. 215-229.

\bibitem{Gelfand2008} I.M. Gelfand, M.M. Kapranov, A.V. Zelevinsky, \textit{
Discriminants, Resultants, and Multidimensional Determinants}, Springer
Science \& Business Media, 2008.

\bibitem{Janson2010} S. Janson, Resultant and discriminant of polynomials,
Lecture notes, \url{ http://www2.math.uu.se/\symbol{126}svante/papers/sjN5.pdf},
2010.

\bibitem{Nusse1997} H.E. Nusse, J.A. Yorke, 
\textit{Dynamics: Numerical Explorations}, Springer Verlag New York Inc, 1997.

\end{thebibliography}
\end{document}